\documentstyle[twocolumn,prl,aps,epsf]{revtex}
\begin{document}

\wideabs{ \draft
\title{Observation of band structure and density of states effects in Co-based magnetic tunnel junctions}
\author{P. LeClair\footnotemark[1], J.T. Kohlhepp, C.H. van de Vin, H. Wieldraaijer, H.J.M. Swagten, and W.J.M. de Jonge}
\address{Department~of~Applied~Physics, Eindhoven~University~of~Technology,
P.O.~Box~513, 5600 MB~~Eindhoven, The~Netherlands}
\author{A. H. Davis and J. M. MacLaren}
\address{Department~of~Physics, Tulane~University, New~Orleans, LA~70118}
\author{R. Jansen and J.S. Moodera}
\address{Francis~Bitter~Magnet~Laboratory, Massachusetts~Institute~of~Technology, Cambridge, Ma~02139}
\maketitle
\begin{abstract}

Utilizing Co/Al$_2$O$_3$/Co magnetic tunnel junctions (MTJs) with Co
electrodes of different crystalline phases, a clear relationship between electrode structure and junction transport properties is presented. For junctions with one fcc(111) textured and one polycrystalline (poly-phase and poly-directional) Co electrode, a strong asymmetry is observed in the magnetotransport properties, while when both electrodes are polycrystalline the magnetotransport is essentially symmetric. These observations are successfully explained within a model based on ballistic tunneling between the calculated band structures (DOS) of fcc-Co and hcp-Co. 

\end{abstract}

  }

\footnotetext[1]{Corresponding author. Electronic mail: pleclair@phys.tue.nl}

%%% introduction and motivation %%%

One of the most basic and essential aspects of magnetic tunnel junctions~\cite{moodera95} (MTJs) is the role of the electronic structure in determining MTJ properties. In general, the sensitivity of tunnel current to the electronic structure of the tunneling electrodes has been well established theoretically for some time ~\cite{appelbaum-brinkman2,levy_review,tsymbal00}. Experimentally, there have been few conclusive observations of the electronic structure of a {\it normal} ({\it i.e.,} non-superconducting) metal in tunneling experiments, despite extensive efforts~\cite{wolf,jaklevic73,rowell69}. Spin polarized tunneling in superconductor-insulator-ferromagnet junctions~\cite{spt}, for example, can only reveal the spin polarization of the participating density of states (at the Fermi level) of electrons tunneling from the ferromagnet. More recently, De Teresa {\it et al}~\cite{fert_lsmo} have used half-metallic La$_{0.7}$Sr$_{0.3}$MnO$_3$ (LSMO) in Co/SrTiO$_3$/LSMO junctions to analyze the sign and magnitude of the Co spin polarization as a function of bias, pointing out the importance of participating tunneling states. Yuasa {\it et al}~\cite{yuasa00} have also recently shown a dependence of the tunneling spin polarization on crystallographic orientation for epitaxial Fe electrodes. Still, few examples of a clear experimental correlation between the conductance properties and the electronic structure of normal metal electrodes have been identified, particularly for magnetic junctions, .

The difficulties in observing band structure or density of states (DOS) effects in normal metal junctions (or MTJs) are numerous. One limitation is that one can only hope to see band or DOS features for those bands that contribute to the tunnel conductance. These bands and the DOS they generate are (for the Al$_2$O$_3$ barriers utilized) believed to be limited to highly dispersive bands whose states at finite $k$ are $s$-hybridized~\cite{levy_review,tsymbal00,wolf,spt,butler.mtj}. A second complication is the presence of many conductance contributions, such as normal elastic tunneling, and inelastic excitations (phonons, magnons), which may obscure electronic structure features, but which may be essentially eliminated with careful analysis~\cite{wolf}. A third complication is the extreme difficulty in theoretical analysis of these structures, particularly MTJs. A last, and perhaps most fatal complication is the need for tunneling electrodes with a {\it known physical and electronic structure which can be modulated} in order to convincingly compare theory and experiment. 

In this Letter, we prepare Co/Al$_2$O$_3$/Co MTJs with differently textured Co layers, {\it viz.}, highly-textured fcc(111)-Co and polycrystalline-Co, in order to realize a model system in which to study the relation between electronic and physical structure and MTJ properties. Using only Co, but in different crystalline phases (and hence with different electronic structure) crucially allows us to focus {\it only} on changes in the electronic and physical structure. By a careful analysis of the conductance-voltage and magnetoresistance-voltage characteristics, in combination with $^{59}$Co Nuclear Magnetic Resonance (NMR) {\it to directly investigate local physical structure}, we show experimentally that junctions with electrodes of two different crystalline phases show obviously asymmetric transport characteristics. The observed asymmetries can be qualitatively, and to some extent, quantitatively, explained by a model developed by the authors~\cite{davis.thesis,temp.dep.davis,bias.dep.davis} based on ballistic tunneling between the calculated band structures (and resulting DOS) of fcc-Co and hcp-Co. Furthermore, these experimental and theoretical results may explain several related observations of similar conductance asymmetries~\cite{rowell69,zhang.bd,moodera.iets,philips.asym}. 

%%% experimental stuff %%%

Ferromagnetic tunnel junctions were prepared by UHV dc/rf
magnetron sputtering (base pressure $<$5$\times$10$^{-10}$ mbar); the details of this fabrication process have been
described elsewhere~\cite{dusting.prl,conductance.apl}. The TMR
structures used consisted of : Si(001)/SiO$_2$/{\it buffer}/Co {\it d}$_{\mbox{\scriptsize
Co}}$/Al$_2$O$_3$/Co 150\AA/Al 30\AA\, post-annealed in a magnetic
field at 200$^{\circ}$C. For these experiments, the buffer consisted of either Ta 50\AA/Co 70\AA/FeMn 100\AA\ (``FeMn-based buffer") or
only Ta 50\AA (``Ta buffer"). Junction resistances or dynamic
resistances ($\frac{dV}{dI} \equiv G^{-1}$) were measured using standard {\it dc} or {\it ac} lock-in techniques. {\it In all cases, the bottom electrode was biased positively for V$>$0.} 

In order to prove that Co layers grown on
FeMn-based and Ta-based buffers have a different physical structure,
we have performed $^{59}$Co Nuclear Magnetic Resonance (NMR)
measurements on separately grown Si/SiO$_2$/Ta 50\AA/Co {\it
d}$_{\mbox{\scriptsize Co}}$/Al$_2$O$_3$ (Ta-based), Si/SiO$_2$/Ta
50\AA/Cu 70\AA/FeMn 100\AA/Co {\it d}$_{\mbox{\scriptsize
Co}}$/Al$_2$O$_3$ (FeMn-based), and Al$_2$O$_3$/Co 150\AA\ (to
represent the top electrode) structures, sputtered under identical
conditions.~\cite{cu_co} The NMR experiments were performed at 1.5 K in zero applied field, as described in ~\cite{nmr.analysis2}. NMR measurements
give information on the distribution hyperfine fields
at the Co nuclei, from which it is possible to distinguish Co atoms in different structural environments ({\it e.g.}, fcc, hcp, bcc) and to determine their relative amounts~\cite{nmr.analysis2,nmr.analysis}. 

Fig.\ 1 shows the resulting $^{59}$Co NMR intensity as a function
of frequency for FeMn-based structures (Fig. 1a) with {\it
d}$_{\mbox{\scriptsize Co}}$=35, 50, 80\AA, Al$_2$O$_3$/Co/Al structures with {\it d}$_{\mbox{\scriptsize Co}}$=80\AA, and for a Ta-based structures with {\it d}$_{\mbox{\scriptsize Co}}$=50\AA\ (Fig. 1b).
Vertical lines indicate the frequencies for bulk hcp (magnetization $\|$ or
$\perp$ to the c-axis) and fcc Co. Focusing on the FeMn-based
structures, Fig.\ 1a, a strong line is observed at the frequency
corresponding to fcc-Co for {\it d}$_{\mbox{\scriptsize
Co}}=$ 50\AA\ and 80\AA, with little intensity at the hcp-Co
positions or intervening regions (corresponding to stacking
faults~\cite{nmr.analysis2,nmr.analysis}). For the 35\AA\ sample,
the dominant line is still fcc-Co, but a considerable intensity
exists at the hcp-Co and stacking fault positions. However, from
35-80\AA\ the stacking fault intensity does not increase. From
this, we may conclude~\cite{nmr.analysis2,nmr.analysis} that Co
growth begins with a considerable amount of stacking faults, but
then grows in a more ordered fcc structure, with presumably the
purest fcc-Co present at the uppermost surface. Thus, even the
35\AA\ Co layer has predominantly fcc-Co stacking
for the topmost layers adjacent the barrier. Combining the NMR data with X-ray
diffraction measurements indicates that the Co layers with FeMn-based buffers are (111) textured fcc-Co.

Fig.\ 1b shows the NMR intensity spectrum for representative
Ta-based and Al$_2$O$_3$ based samples. No significant Co thickness dependence was observed,
indicating the growth mode is the same for all thicknesses. For
either the Ta-based samples or those grown on Al$_2$O$_3$, intensity is observed for both hcp positions as well as the fcc position and all intermediate stacking fault positions. From this, we may
conclude that Co grows on Ta or Al$_2$O$_3$ in a random,
poly-crystalline and poly-phase manner~\cite{nmr.analysis2,nmr.analysis} (which we refer to as ``poly-Co" for simplicity), in contrast to the fcc(111)-textured Co in FeMn-based structures. Thus, the FeMn-based and Ta-based MTJs prepared for this study may be clearly identified as {\it asymmetric} fcc(111)-Co/Al$_2$O$_3$/poly-Co and {\it symmetric} poly-Co/Al$_2$O$_3$/poly-Co junctions respectively. We must caution that although the FeMn-based junctions produce {\it strongly fcc(111)-textured} bottom electrodes, they contain a fraction of Co atoms which remain in a non-fcc environment.

% G(V) part%

One may anticipate that these structural differences lead to differences in band structure and DOS, and hence, differences in transport properties. From symmetry, one expects that a physically asymmetric junction results in transport which is asymmetric with respect to bias. Fig.\ 2 shows the normalized conductance-voltage ($\frac{dI}{dV}(V) \equiv G(V)$) characteristics for parallel Co magnetizations at 5K for two
fcc(111)-Co {\it d}$_{\mbox{\scriptsize Co}}$/Al$_2$O$_3$/poly-Co junctions with {\it d}$_{\mbox{\scriptsize Co}}$ = 35, 50\AA\, and a poly-Co/Al$_2$O$_3$/poly-Co junction. The fcc(111)-Co/Al$_2$O$_3$/poly-Co junctions show an obvious conductance asymmetry, with most notably a local minimum at $\sim$-0.25V and a slight ``shoulder" at the same positive voltage, only for parallel magnetizations. Parenthetically, we
note that the asymmetry was also present for intentionally over- or under-oxidized barriers. We will return to the explanation of these unusual features shortly. The poly-Co/Al$_2$O$_3$/poly-Co
junctions show almost perfectly symmetric behavior for both magnetization orientations, with a parabolic background and a low-voltage linear
contribution, with no apparent thickness dependence. For both types of junctions, the former
parabolic portion can be understood in terms of regular elastic
tunneling~\cite{wolf}, whereas the linear contribution is
consistent with magnon-assisted tunneling~\cite{zhang.bd}.
However, the local-minimum and shoulder observed for fcc(111)-Co/Al$_2$O$_3$/poly-Co junctions cannot be explained by either of these mechanisms. The striking difference between junctions where only the crystalline phase of the bottom electrode was intentionally altered
suggests that the electronic structure and spin-dependent DOS of the fcc-Co structure must be responsible.

% bias dependence asymmetry part%

If the DOS and band structure differences between the
two Co electrodes are responsible, one may expect the
bias dependence of the tunnel magnetoresistance (TMR) to be altered as well. Although the TMR was generally lower for poly-Co/Al$_2$O$_3$/poly-Co structures, this criterion is very susceptible to slight differences in preparation conditions. The normalized {\it dynamic} TMR-voltage ($\Delta G/G_a$-V) characteristics, though generally nearly identical for many junctions~\cite{conductance.apl}, suffer from a strong contribution by inelastic excitations~\cite{zhang.bd} which may mask the underlying electronic structure effects. Since we expect that to first order any inelastic excitations give a contribution {\it symmetric} in applied bias, by plotting the {\it odd} part of the dynamic MR-voltage behavior we may be primarily sensitive to only electronic structure (or self-energy) effects~\cite{wolf}. We emphasize that this sort of analysis is {\it crucial} for comparison with theory, effectively eliminating those contributions which are difficult to incorporate theoretically. Fig.\ 3a shows the odd portion of $\Delta G/G_a$ ({\it i.e.}, $\Delta G/G_a$ (V$>$0) - $\Delta G/G_a$ (V$<$0)) as a function of voltage, a measure of the asymmetry in the bias dependence of the {\it dynamic} MR, $\Delta G/G_a$~\cite{note.bd}. For
poly-Co/Al$_2$O$_3$/poly-Co junctions, almost no asymmetry is present, as expected for nominally identical electrodes. However, for fcc(111)-Co/Al$_2$O$_3$/poly-Co junctions, there is an obvious strong minimum, corresponding to the same voltage where the local minimum and shoulder features were seen in the dI/dV(V) curves. This further suggests that the electrode DOS and band structure may play a key role. Since the relative dynamic MR more strongly {\it decreases} at the position of the
shoulder and local minimum, one expects that either the minority
DOS increases, or the majority DOS decreases. As we will show, the dip and shoulder features found in G(V) can be accounted for theoretically by a combination of band structure and DOS arguments, incorporated into a ballistic model of tunneling~\cite{davis.thesis,temp.dep.davis,bias.dep.davis}.

%theory part comes%

The DOS and band structures used were calculated with the layer-KKR method~\cite{DOS.calc}. This approach uses spherical potentials calculated self-consistently within the local spin density and atomic spheres approximations. Band structure is determined from the scattering properties of the atomic layers, and the DOS is determined from the imaginary part of the Green's functions~\cite{DOS.calc}. For simplicity, we use hcp bands to model the poly-Co electrode, since we are looking primarily at features of the fcc DOS. Although it has been previously shown that tunneling is extremely interface sensitive~\cite{dusting.prl,cr.dusting}, we expect that the Co
band structure and DOS at the barrier interface, though certainly
altered, will strongly resemble the bulk, at least with regard to the
general features one may observe~\cite{note.interface_states}. We first observe that the underlying tendency for dI/dV to dip in fcc(111)-Co/Al$_2$O$_3$/poly-Co structures can be explained in terms of the DOS for states with $s$-character above and below E$_F$ in the fcc electrode. The calculated DOS for fcc-Co~\cite{davis.thesis} shows that the total DOS is negatively polarized at E$_F$, but that the $s$-DOS is positively polarized. This suggests that states with $s$-character dominate the tunneling with these barriers~\cite{tsymbal00,spt,butler.mtj}. Calculations by the authors show that the most dispersive bands, corresponding to those which have previously been identified as the source of the tunneling conductance, exhibit significant $s$-hybridization. Since these are hybridized states, every contribution to the $s$-character of the band is accompanied by a contribution to the $d$-character from the same state (and to a smaller extent, the $p$-character). In each case, for Fe, Co, and Ni, there is only a single band with significant $s$-character near E$_F$. Therefore, the $s$-DOS emerges as a good marker for the dynamic behavior of the conductance (dI/dV) in this bias range. 

The fcc-Co $s$-DOS is known to exhibit two sharp peaks at about 0.4 eV above {\it and} below E$_F$~\cite{davis.thesis}; DOS peaks imply localized states that have a reduced tunneling
probability~\cite{spt,butler.mtj,davis.thesis}. For the fcc(111)-Co/Al$_2$O$_3$/poly-Co structures, note that in the experimental configuration electrons flow from hcp to fcc bands for positive bias, and fcc to hcp bands for negative bias. For negative bias, tunneling {\it into the poly-Co electrode}, biasing now includes {\it occupied} states
from {\it below} E$_F$ in the fcc electrode. Since the fcc states
below E$_F$ are increasingly localized, dI/dV is suppressed,
resulting in a dI/dV dip. For tunneling {\it into the fcc
electrode} at positive bias, we expect the same tendency of dI/dV to decrease, because now localized states {\it above} E$_F$ are participating. However, for positive bias, an additional effect enters the picture. In the fcc-Co band structure, there is an unoccupied, but very dispersive, minority band which begins just above E$_F$~\cite{davis.thesis}. Since it is {\it above} E$_F$, it only participates for tunneling {\it into} the fcc-Co electrode {\it from} the poly-Co electrode (positive bias). This highly dispersive band contributes spin down states, thus decreasing the MR, but augmenting dI/dV, since it leads to an overall increase in available states. The emergence of this band thus partially suppresses the tendency of dI/dV to dip. As a result, the shoulder observed for positive bias in dI/dV looks suspiciously like a dip which never fully develops. Since this band participates only for positive bias, it suppresses the MR only for positive bias, leading to a minimum in the odd part of the dynamic MR. Interestingly, earlier work with Cu dusted MTJs~\cite{conductance.apl,ru.dusting} showed that dusting of the fcc electrode suppresses the dip in dI/dV, which in hindsight may further indicate that the dips are fcc-Co specific features.

Fig.\ 3b shows the odd part of the dynamic TMR, as in Fig.\ 3a, {\it calculated} for a fcc-Co/I/hcp-Co MTJ using the model described in Refs.\ ~\cite{temp.dep.davis,bias.dep.davis}. For the poly-Co/I/fcc-Co structure, we first note the small peak at small biases, related to the aforementioned minority fcc-Co band beginning just above E$_F$. When this minority fcc-Co band is artificially shifted to begin {\it below} E$_F$ instead of {\it above} E$_F$, the small peak disappears, clearly indicating that the minority band beginning just above E$_F$ in fcc-Co is responsible for this feature. Thus, there is an overall tendency for the odd part of the dynamic TMR to increase at low biases (the majority current rises more rapidly than the minority current), but when the minority band just above E$_F$ emerges for slightly larger biases, there is a strong increase in minority tunneling states, and the odd part of the dynamic TMR decreases. 
This again indicates that the small peak in the odd part of the TMR is {\it directly related to the unoccupied minority band of the fcc-Co structure}, and the localized part of the $s$-DOS is responsible for the underlying overall tendency of dI/dV to dip. We note parenthetically that the calculated bias dependence for hcp-Co/I/hcp-Co junctions is
completely symmetric, and the odd part of the dynamic TMR is thus identically zero.  Finally, we note that these results may explain several similar observations in literature~\cite{rowell69,moodera.iets,philips.asym}. 

In conclusion, we have shown for the first time a clear relation between the
electronic and physical structure of the magnetic electrodes and the
magneto-transport properties of MTJs. When
both electrodes are poly-Co, only residual asymmetry is observed, whereas when one electrode is textured fcc(111)-Co,
strongly asymmetric characteristics are observed. The asymmetric
magneto-transport properties may be understood by considering the
itinerant electron bands in the two crystalline phases of Co, as
determined from first-principles calculations, and incorporated into a
model of ballistic tunneling. 

P. LeClair is supported by the Netherlands technology foundation STW, and work at Tulane was supported by DARPA award MDA 972-97-1-003.

%\bibliographystyle{prsty}
%\bibliography{refs}

\begin{thebibliography}{10}
\bibitem{cu_co}
In this case, a Cu layer was used rather than Co as a polycrystalline seed
  layer for the FeMn layer, but this has no effect on the resulting texture.
\bibitem{note.interface_states}
We expect Co/Al$_2$O$_3$ interface states are localized, and hence, not
  contribute significantly to the tunnel current.
\end{thebibliography}

\begin{figure}[htb]
\begin{center}
\hspace{0cm}\epsfxsize=0.95\columnwidth \epsfbox{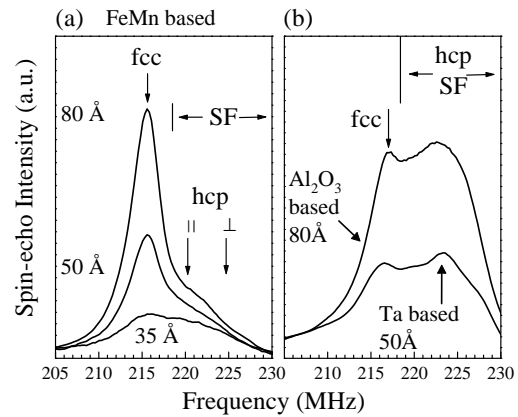}
\caption[]{$^{59}$Co NMR spectra at 1.5K in zero applied field for (a) FeMn-based structures with varying {\it d}$_{\mbox{\scriptsize Co}}$ (see text), and (b) Ta/Co/Al$_2$O$_3$ and Al$_2$O$_3$/Co/Al structures.
 } \label{NMR}
\end{center}
\end{figure}

\begin{figure}[htb]
\begin{center}
\hspace{0cm}\epsfxsize=0.95\columnwidth \epsfbox{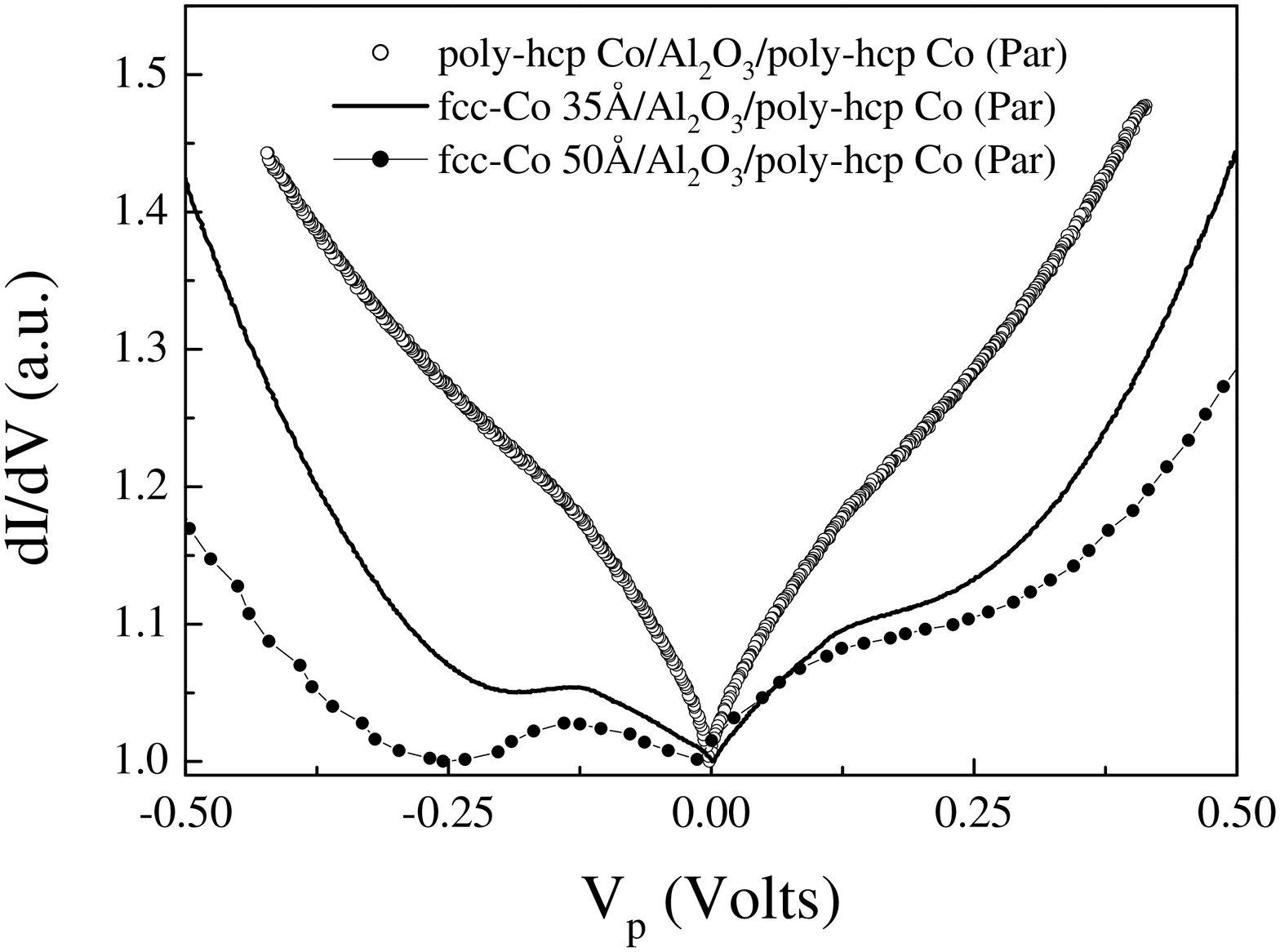}
\caption[]{Conductance-voltage characteristics for
fcc(111)-Co 35\AA\, 50\AA\ /Al$_2$O$_3$/poly-Co junctions, and a poly-Co/Al$_2$O$_3$/poly-Co junction, all for parallel magnetization alignments at 5K.} \label{GV}
\end{center}
\end{figure}

\begin{figure}[htb]
\begin{center}
\hspace{0cm}\epsfxsize=0.95\columnwidth \epsfbox{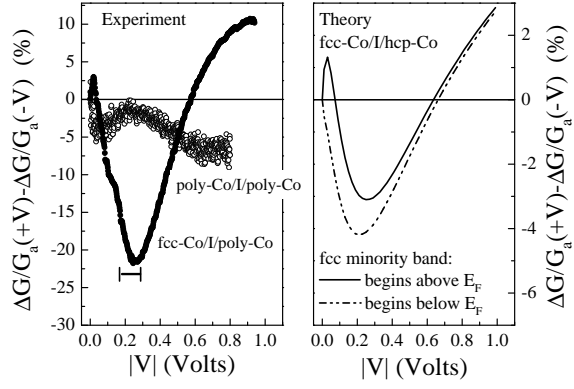}
\caption[]{Odd part of $\Delta G/G_{a}$ vs. voltage (a)
experimental data at 5K for a fcc(111)-Co/Al$_2$O$_3$/poly hcp-Co junction 
and a poly-Co/Al$_2$O$_3$/poly hcp-Co junction. Approximate position of
conductance local minimum and shoulder is denoted with the horizontal bar. (b) same curve calculated within the ballistic tunneling model for a fcc-Co/I/hcp-Co junction (see text).} \label{bias_dep}
\end{center}
\end{figure}

\end{document}